\documentclass[twocolumn,showpacs,preprintnumbers,amsmath,amssymb,aps,prl,superscriptaddress]{revtex4-1}
\usepackage{color}
\usepackage{graphicx}
\usepackage{textcomp} 
\usepackage{subeqnarray}
\usepackage[caption=false]{subfig} 
\usepackage{ulem}
\newcommand\be{\begin{equation}}
\newcommand\bea{\begin{eqnarray} \nonumber }
\newcommand\ee{\end{equation}}
\newcommand\eea{\end{eqnarray}}

\begin{document}
\title{Unravelling the trading invariance hypothesis}

\author{Michael Benzaquen}
\affiliation{Capital Fund Management, 23 rue de l'Universit\'e, 75007 Paris}
\author{Jonathan  Donier}
\affiliation{Capital Fund Management, 23 rue de l'Universit\'e, 75007 Paris}
\author{Jean-Philippe Bouchaud}
\affiliation{Capital Fund Management, 23 rue de l'Universit\'e, 75007 Paris}
\affiliation{CFM-Imperial Institute of Quantitative Finance, Department of Mathematics, Imperial College, 
180 Queen's Gate, London SW7 2RH}

\medskip


\begin{abstract}
We confirm and substantially extend the recent empirical result of Andersen et al. \cite{Andersen2015}, where it is shown that the amount of risk $W$ exchanged in the E-mini S\&P futures market (i.e. price times volume times volatility) scales like the 3/2 power of the number of trades $N$. We show that this 3/2-law holds very precisely across 12 futures contracts and 300 single US stocks, and across a wide range of time scales. However, we find that the ``trading invariant'' $I=W/N^{3/2}$ proposed by Kyle and Obizhaeva is in fact quite different for different contracts, in particular between futures and single stocks. Our analysis 
suggests  $I/{\cal C}$ as a more natural candidate, where $\cal C$ is the average spread cost of a trade, defined as the average of the trade size times the bid-ask spread. 
We also establish two more complex scaling laws for the volatility $\sigma$ and the traded volume $V$ as a function of $N$, that reveal the existence of a characteristic number of trades $N_0$ above which the expected behaviour $\sigma \sim \sqrt{N}$ and $V \sim N$ hold, but below which strong deviations appear, induced by the size of the~tick. 
\end{abstract}
\maketitle

\section{Introduction}

Understanding the dynamics of financial markets is of obvious importance for the financial industry, but also for decision makers, central bankers and regulators. It is also a formidable
intellectual challenge that has attracted the interest of many academic luminaries, with perhaps Benoit Mandelbrot as a legendary figure.  He was the first to propose
the idea of {\it scaling} in this context \cite{Mandelbrot1997}, a concept that in fact blossomed in statistical physics before getting acceptance in economics and finance (for a review, see \cite{Gabaix2009}). 
In the last twenty years, many interesting 
scaling laws have been reported, concerning different aspects of price and volatility dynamics. One particular question that has been the focus of many studies is the relation between volatility and trading activity, measured as the number of trades and/or the volume traded (see e.g. {\cite{Clark1973, Tauchen1983, Jones1994, Bollerslev1999,Ane2000,Liesenfeld2001,Tauchen1983,Engle1996}} and more recently \cite{Zumbach2004,Eisler2006,Wyart2008}). Revisiting these results, Kyle and Obizhaeva (KO) recently proposed a bold but inspiring hypothesis, coined as the \textit{trading invariance principle}. 

Their original idea primarily relies on {\it dimensional analysis}, which is very common in physics and states that any ``law'' relating different 
observables must express one particular dimensionless (or unit-less) combination of these observables as a function of one or several other such 
dimensionless combinations. The simplest example might be the ideal gas law, that
amounts to realizing that pressure $p$ times volume $v$ has the dimension of an energy. Hence $pv$ must be divided by the thermal energy $RT$ of a mole of gas to 
yield a dimensionless combination. The right-hand 
side of the equation must be a function of other dimensionless variables, but in the case of non-interacting point-like particles, there is none -- hence the only possibility is $pv/RT = \text{cst}$. 
Deviations from the ideal gas law are only possible because of the finite radius of the molecules, or the strength of their interaction energy, that allows one to create other dimensionless  
combinations (and correspondingly new interesting phenomena such as the liquid-gas transition!).

In the search of an ``ideal market law'' for stocks, several possible observable quantities that characterize the trading activity come to mind: the total market capitalisation $M$ (in dollars), the share price $P$ (in dollars per share), the square volatility $\sigma^2$ (in $\%^2$ per day), the amount traded $V$ (shares per day), and the volume of individual ``bets'' $Q$ (in shares) \footnote{A bet is the ensemble of trades that originate from a single trading decision. It is alternatively called a ``metaorder'' in the literature.}. Other, more microstructural quantities might come into play, such as the difference between the best bid and best offer price, called the spread $S$ (in dollars per share), the tick size $s$ (in dollars) that fixes the smallest possible price change, the lot size $\ell$ (in shares) that fixes the smallest amount of exchanged shares, the average volume available at the 
best quotes, and perhaps other quantities as well. 

Kyle and Obizhaeva further postulate the existence of a universal invariant $I$ in dollars, that they interpret as the average ``cost'' of a single bet, and keep only $P, \sigma^2, V$ and $Q$ as relevant variables. Dimensional analysis then immediately leads to the following relation:
\be\label{general}
\frac{PQ}{I} = f\left(\frac{Q \sigma^2}{V}\right)
\ee
where $f$ is a certain function that cannot be determined on the basis of dimensional analysis only. At this point, Kyle and Obizhaeva \cite{KyleMM2016} invoke the Modigliani-Miller theorem and argues that capital restructuring between debt and equity should keep $P \times \sigma$ constant, while not affecting the other variables. This suggests that $f(x) \sim x^{-1/2}$, finally leading to the KO \textit{trading invariance principle}:  \footnote{Up to a redefinition of $I$, one can always set $f(x)=x^{-1/2}$ without any numerical prefactor.}
\begin{equation}\label{MMI}
I = \frac{P \sigma Q^{3/2}}{V^{1/2}}:=\frac{W}{N^{3/2}} \ ,
\end{equation}
where $W:=P V \sigma$ is a measure of {exchanged risk}  (precisely the dollar amount of risk traded per day), also referred to as \textit{trading activity} by Andersen and coauthors~\cite{Andersen2015},  and $N:=V/Q$ represents the number of bets per day. 

This simple scaling relation was empirically confirmed by KO using portfolio transition data \cite{Kyle2010}. Portfolio transitions correspond to rebalancing decisions by institutional investors, that are then executed by brokers who collated the corresponding data. However, these trades only reflects part of the 
market activity, and it is furthermore not obvious that these portfolio transitions can be associated with elementary bets. Andersen \textit{et al.} \cite{Andersen2015} reformulated KO's {invariance principle} in a way that can be tested on public trade-by-trade data. Their analysis on the E-mini S\&P 500 futures contract showed that Eq. (\ref{MMI}) holds
remarkably well at the single-trade scale. In this context, $Q$ denotes the average volume of trades and $N$ is the total number of trades within some time
interval $\tau$ ($1$ minute in their analysis).
Because the activity of the market has significant intraday variability, notably marked by the switching from Asian to European and American trading hours, $N$ typically varies over almost two decades, indeed allowing one to test the scaling relation $W \sim N^{3/2}$ quite convincingly (see Fig. 1 below). 

Such a remarkable empirical result, and its purported universal status, clearly cries for further scrutiny and interpretation. Indeed, the idea of Andersen \textit{et al.} \cite{Andersen2015} that the trading invariance hypothesis can be downscaled from bets to trades is far from obvious. Although bets are made of a collection of successive trades, the way in which bets are shredded into trades  significantly depends on the investor and the market \cite{KyleTuzun2016,Bae2014}. The goal of this paper is to dissect the trading invariance hypothesis on a wide range of futures contracts and individual stocks. Equation (\ref{MMI}) can actually be interpreted in different ways, depending on the degree of universality attached to its validity:
\begin{enumerate}
\item {\it No universality}: The scaling relation $W \sim N^{3/2}$ (the ``3/2-law'' henceforth) holds for some contracts and some time intervals $\tau$ (over which $W$ and $N$ are computed). 
In the cases where the scaling law holds, the prefactor $I$ has a non-universal value (that depends on the contract and/or on $\tau$).
\item {\it Weak universality}: The 3/2-law holds for all contracts and some (possibly all) time intervals $\tau$, but with a non-universal value of $I$.
\item {\it Strong universality}: The 3/2-law holds for all contracts and all time intervals $\tau$, with a universal value of $I$, independent of $\tau$ and of the contract type.
\end{enumerate}
The last case might in fact be too strong: it would already be a remarkable result that $I$ only depends on the contract type (say stocks) and on the geographical zone (say the U.S.). In fact, 
from general considerations it would be very strange that $I$ (in dollars) is completely universal, for one thing because the value of the dollar itself is time dependent. 
As we will show in detail below, our results favor the second interpretation of ``weak universality'' where the 3/2-law holds for all contracts, and all time 
intervals $\tau$. However, the value of $I$ itself varies significantly, both within the universe of US stocks and among the different futures contracts. 
Furthermore, the separate analysis of the scaling of $\sigma$ vs. $N$ on the one hand and $V$ vs. $N$ on the other (the product of the two essentially leading to the 3/2-law) 
reveals a surprisingly rich and universal behaviour, and suggests that $W \sim N^{3/2}$ might only be an approximation. 

The outline of the paper is as follows. In section 1, we replicate and confirm Andersen \textit{et al.}'s results on E-mini S\&P 500 futures contract \cite{Andersen2015} and extend 
them to eleven  other futures contracts. We show that the 3/2-law does hold both across time and across contracts, but that the average value of $I$ (and the whole distribution of $I$, for that matter) clearly depends on the considered contract. In section 2,  we confirm the 3/2-law across a pool of 300 US stocks  and show that microstructure effects play a much more important role than in the case of futures contracts. 
In section 3 we propose a unifying picture that decomposes the 3/2-law into two more fundamental scaling laws, that allow us to rescale all futures contracts and all time scales onto two universal master curves.  
Similar to the deviations away from the ideal gas law {example} alluded to above, our results suggest that  
additional microstructural variables must be involved in the search of a relation generalizing Eq. (\ref{general}), where the bid-ask spread and the tick size, among other things, should play an important role -- like the 
molecular size in the ideal gas analogy. {In section 4, we suggest an alternative and more natural definition for trading invariant that accounts some of the microstructural details mentioned above.}

\section{1. Futures contracts} 
\label{secFutures}

We have analysed tick by tick data for the best bid and offer of twelve different futures contracts spanning over three years, from January 2012 to December 2014 (see Tab.~\ref{SummaryFut}). We consider front month contracts only, among which { three} index futures, { four} energy futures, two agriculture futures, one bond future, one FX future and one metals future. All contracts are traded basically twenty-four hours a day, five days a week, on the CME, NYBOT, NYMEX, ECBOT, COMEX,  ICUS and IPE electronic platforms. Three trading regimes can be distinguished corresponding respectively to Asian, European and American regular trading hours. At variance with the analysis of Andersen \textit{et al.} \cite{Andersen2015}, we do not discard any time intervals from our study since we found that doing so did not significantly change the results.

\begin{figure}[t!]
  \centering 
  \includegraphics[scale=0.55]{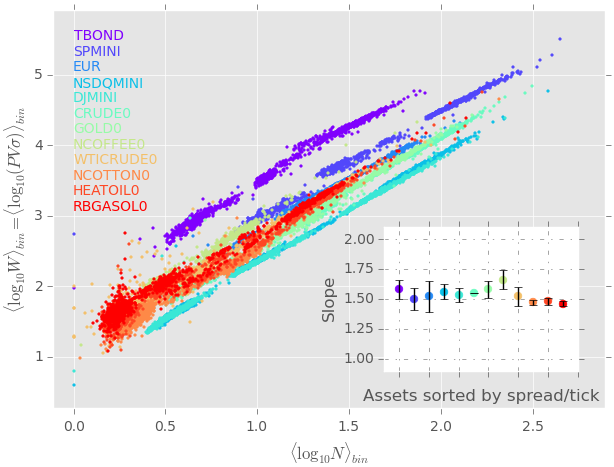}
 \caption{Scatter plot of { $\langle \log_{10} W \rangle_{\text{bin}}$ \textit{vs.} $\langle \log_{10} N \rangle_{\text{bin}}$ }for twelve different futures contract sorted by spread over tick values from cold (large ticks) to warm colours (small ticks). The inset shows the slopes $\alpha$ obtained from linear regression of the data, which are all clustered around $3/2$. Spread over tick values as well as the slopes $\alpha$ are provided in  Tab.~\ref{SummaryFut}.}
\label{RegrFut}
\end{figure}

For each contract, we group the trades by market time stamp, under the assumption that simultaneous trades correspond to a market order originating from a single participant.
We then compute trading volume $V$, number of trades $N$, average trade size $Q = V/N$ and average price $P$ within each one minute bin ($\tau=1$ min). We also compute the volatility $\sigma$, from the average of ten second squared-returns. At variance with Andersen and coauthors \cite{Andersen2015}, we do not annualize our volatilities.  Average values of these quantities for $\tau=1$ min, as well as average volume at the bid and the ask and average spread, are provided in Tab.~\ref{SummaryFut}.  Note that throughout the paper we will elicit power-laws by considering linear regression of log quantities (for example $\log W$ vs. $\log N$). Consistent with this procedure, averages shall be defined with respect to the log-transform, and we will write { ${\langle X \rangle }:= \exp\left[ \mathbb E (\log X) \right]$.} 

\begin{figure}[t!]
  \centering 
  \includegraphics[scale=0.54]{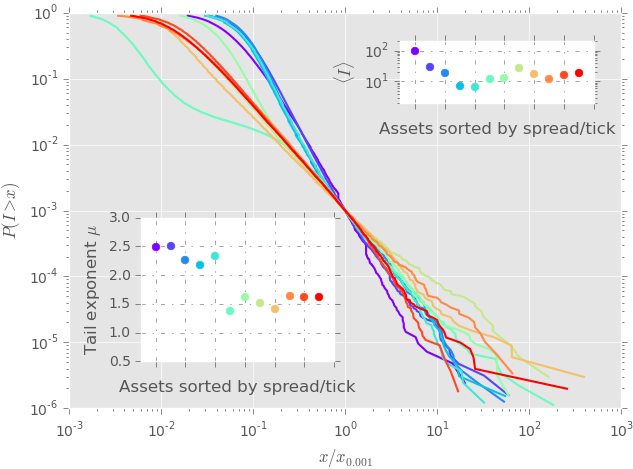}
 \caption{Rescaled complementary cumulative distribution function of $I=WN^{-3/2}$ for twelve different futures contracts sorted by spread over tick values from cold (large ticks) to warm colours (small ticks). The insets show the average values of $I$ (in dollars) and the tail exponents computed according to the Hill estimator  with a cutoff at $P(I>x) = 10^{-2}$. Spread over tick values as well as the average values and tail exponents are provided in Tab.~\ref{SummaryFut}.}
\label{ccdfFut}
\end{figure}

Following the method of Andersen to test the intraday trading invariance hypothesis, we first  average over all days  the logarithm of the aforementioned quantities, for each fixed one minute bin. The latter averaging operator shall be noted $\langle\, . \,\rangle_{\text{bin}}$. Note that taking the logarithm prior to averaging dampens the influence of outliers and leads to a robust estimate of the ``typical value'' of these quantities. The linear regression of $\langle \log W\rangle _{\text{bin}}$ \textit{vs.} $\langle \log N \rangle_{\text{bin}}$ is displayed in Fig.~\ref{RegrFut} and Tab.~\ref{SummaryFut}, and indeed confirms the 3/2-law for all twelve contracts independently. However, the conjecture that the quantity $I=WN^{-3/2}$ -- which visually corresponds to the $y$-intercept of the linear regressions shown in Fig.~\ref{RegrFut} -- is invariant across different contracts is clearly rejected  (see Tab.~\ref{SummaryFut}). 
The top right inset of Fig.~\ref{ccdfFut} displays $\langle I \rangle$ for the twelve futures contracts sorted by spread over tick, and shows that $\langle I \rangle$ varies by a factor $\gtrsim$\,$10$ across different contracts. 
 
\begin{figure}[b!]
  \centering 
  \includegraphics[scale=0.54]{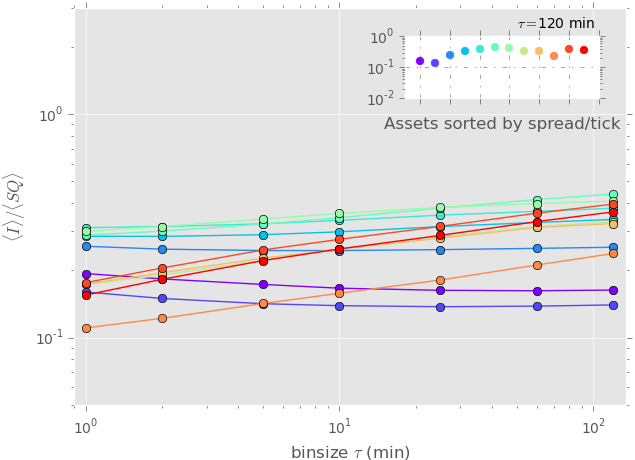}
\caption{Plot of $\langle I\rangle /\langle SQ \rangle$ where $S$ denotes the spread (in dollars per share) and $Q$ denotes the trade size (in shares) computed at the daily scale as a function of bin size $\tau$ (in minutes) for twelve different futures contracts. Note that this ratio is nearly {$\tau$-independent}. The inset shows  $\langle I\rangle /\langle SQ \rangle$ at $\tau = $~120~min sorted by spread over tick values, which is now constant to within a factor $3$ (compare to the top right inset of Fig.~\ref{ccdfFut}).}
\label{NewFig}
\end{figure}

For robustness, we checked that the above results also stand on sub-intervals of one year of the full period 2012-2014. In particular we observe that the variations of $\langle I \rangle$ across contracts (more than a factor 10) are much larger  than the variations from one year to the next for a given contract (around $\sim 20 \%$). The role of the bin size $\tau$ is also very interesting. Averaging over one, five and ten minute bins across days shows consistent results. The analysis on longer time
scales (thirty minute, one and two hour bins) however shows a {slight but} systematic underestimation of the predicted 3/2 slope of $\langle \log W\rangle_{\text{bin}} $ versus $\langle \log N \rangle_{\text{bin}}$ which disappears when the volatility estimator based
on ten-second squared returns is replaced by the Rogers-Satchell volatility estimator \cite{Rogers1991}, known to be more adequate when the underlying follows a geometric Brownian motion with an unknown drift. Note that the Rogers-Satchell estimator measures zero volatility whenever the open price matches the high/low {\it and} the close price matches the low/high, which are not rare events for small bin sizes. Enforcing that the volatility must be non-vanishing leads to discarding a {substantial} fraction of the data at high frequencies. However, we checked that removing the zero volatility intervals has no material impact on the results. In the following we shall thus consistently use the Rogers-Satchell estimator to compute the volatility. \footnote{We have in fact checked that other estimators based on the open, high, low, close prices lead to very similar results.} The conclusion of our analysis is that the 
3/2-law holds across all futures contracts and across all time intervals $\tau$.  Figure \ref{NewFig} displays a plot of the average $ I $ rescaled by the average trade cost $\cal C$, defined as the product of the spread $S$ (in dollars per share) and the trade size $Q$ (in shares), a choice that will be further motivated in section {4}. At this point one should note that a)  $I/\cal C$ is now a dimensionless quantity of order unity, and b) $I/\cal C$ appears to be significantly more stable across assets than $I$ itself (see Fig.~\ref{NewFig}).

Finally, the trading invariance hypothesis -- in its strongest version -- states that the full probability distribution of $I=W/N^{3/2}$ 
(and not only its average value) should be invariant across time and across contracts. To test this point, we have computed the complementary cumulative distribution function $P(I>x)$ for the twelve futures contracts (see Fig.~\ref{ccdfFut}). For the sake of readability, 
the main plot of Fig.~\ref{ccdfFut} displays these distributions with the $x$-axis rescaled by $x_{0.001}$ defined by $P(I>x_{0.001}) = 10^{-3}$. As one can see, the tail of the distributions are all close to power laws. The tail exponent $\mu$ -- defined as $P(I > x) \sim x^{-\mu}$ -- however varies significantly from $\mu \approx 2.5$ for the larger tick futures to $\mu \approx 1.5$ for the smaller tick futures. Tail exponents were computed using the Hill estimator with a cutoff at $P(I > x) = 10^{-2}$ \footnote{The Hill estimator (1975) allows to compute the tail behaviour of a distribution. Defining the tail exponent $\mu$ as $P(X>x)\sim x^{-\mu}$, one has $\mu = \left[\frac1k {\Sigma}_{i=0}^{k-1}  \log(X_i/X_k)\right]^{-1}$ where $k$ denotes the rank of the cutoff.}.  The values of the tail exponents are provided in Tab.~\ref{SummaryFut}.

The conclusions so far are thus:
\begin{enumerate}
\item We fully confirm the  3/2-law found by Andersen \textit{et al.} \cite{Andersen2015} on the E-mini S\&P futures on one-minute intervals;
\item The  3/2-law holds surprisingly accurately for all contracts and all time intervals;
\item The ``invariant'' $I$ is in fact not universal: both its average value and the shape of its distribution function depends quite significantly on the chosen contract. {However, $I$ for a given contract is to a good approximation $\tau$-independent.}
\end{enumerate}
We now extend our analysis to a much wider sample of single stocks, and find that the above conclusions are indeed vindicated.

\section{2. US Stocks} 
\label{secStocks}

\begin{figure}[b!]
  \centering  
  \includegraphics[scale=0.55]{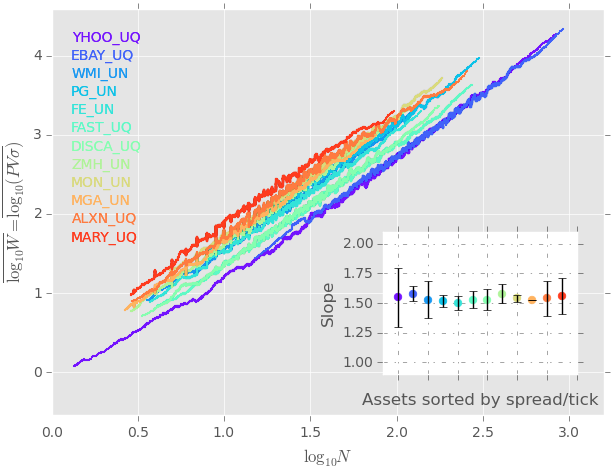}
 \caption{
{ Centred rolling average (window size = 100) of the scatter plot of  $ \log_{10} W $ \textit{vs.} $ \log_{10} N $ for a random subset of twelve different stocks chosen from a pool of three hundred US stocks sorted by spread over tick values from cold (large ticks) to warm colours (small ticks).  The inset shows the slopes obtained from linear regression of the data (before performing the rolling average). Spread over tick values as well as the slopes $\alpha$ obtained from the linear regressions are provided in  Tab.~\ref{SummaryStocks}}.
 }
\label{StocksSlopes}
\end{figure}

Our analysis is conducted on a pool of three hundred US stocks, chosen to be as representative as possible in terms of market capitalisation and tick size.  Note that the large number of assets -- and their diversity -- allows for great statistical significance. 
We consider five-minute bins using trades and quotes data from January 2012 to December 2012, extracted from the primary market of each stock (NYSE/NASDAQ).  We remove auction time intervals, as well {as} thirty minutes after the opening and before the closing of the market, so as to avoid any artefact due to these specific 
trading periods. 
To compute the volatility, we again use the Rogers-Satchell estimator for which only the high, low, open and close prices are needed \cite{Rogers1991}.  The average values of $N$, $Q$, $V$, and $\sigma$ are provided in Tab.~\ref{SummaryStocks} for a random selection of twelve stocks within the pool.

\begin{figure}[t!]
  \centering 
  \includegraphics[scale=0.54]{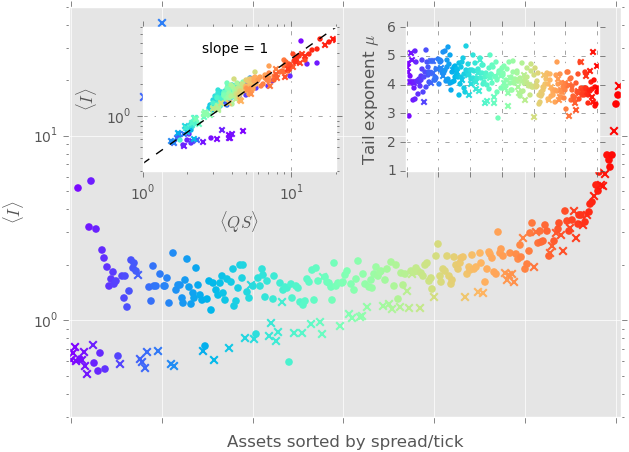}
 \caption{Main plot: average value of $I$ (in dollars) at $\tau=120$ min for three hundred US stocks sorted by spread over tick values from cold (large ticks) to warm colours (small ticks). NASDAQ/NYSE stocks are marked with crosses/filled circles respectively. Top left inset: average value  $\langle I \rangle$ as a function of average trade cost ${\cal C}=\langle SQ \rangle$ {(in dollars)}. Top right inset: tail exponents $\mu$ of the complementary cumulative probability distributions $P(I>x) \sim x^{-\mu}$. Numerical values of $\langle I \rangle$ as well as tail exponents are provided in Tab.~\ref{SummaryStocks} for a random subset of twelve stocks.}
\label{ProbaStocks}
\end{figure}

As in the previous section, we perform a linear regression of $ \log W $ versus $ \log N $ for each stock. {Fig.~\ref{StocksSlopes} is analogous to Fig.~\ref{RegrFut}, only here we do not compute the average of the bins across days as was done in the previous section. This is due to the fact that, unlike futures, the stocks we consider are exclusively traded during American hours and thus lack the ``three-continent" seasonality of the futures. Proceeding as 
suggested by Andersen {\it et al.} for futures would thus significantly reduce the range of possible values of $V, Q, N$ and $\sigma$, thereby degrading the determination of the slopes of the fits. For the sake of readability, Fig.~\ref{StocksSlopes} shows a centred rolling average along $\log N$, with window size of one hundred data points. However, all regressions were performed before the rolling average. For the three hundred stocks, a cross sectional 
determination of the slope yields $\alpha= 1.54 \pm 0.11$, where the  uncertainty here is the root mean square  cross-sectional dispersion. This is again in very good agreement with the prediction $\alpha = 3/2$, {thereby} considerably bolstering the results of the previous section. We also checked that these results hold 
unchanged for lower frequencies, and in particular at daily time scales $\tau=$ 6 hours.

Figure~\ref{ProbaStocks} displays the average values of $I$ (computed using {$\tau=5$ min}) as well as the tail exponents $\mu$ of the complementary cumulative probability distributions for three hundred stocks, sorted by spread over tick from left to right. The typical value of $I$ for the stocks is on average one order of magnitude smaller than for futures -- i.e. the ``bet sizes'' are smaller in dollars on individual stocks than on futures, which is not very surprising. The main plot reveals a striking feature: the appearance of two distinct branches in the larger ticks region. The higher branch presents an intriguing U-shape, somewhat similar to what was observed for futures
contracts, while the lower branch is consistent with a nearly linear dependence on the average spread. Remarkably, the two branches correspond 
chiefly to stocks traded on the NYSE (upper branch) and NASDAQ (lower branch) platforms. For better readability, NASDAQ stocks are represented by crosses while NYSE stocks appear as filled circles. 
Several points could actually explain this difference -- although our understanding of this effect is only partial. For example a non-negligible fraction of the trades on NASDAQ happen within the spread (hidden trades), a particularity that would naturally affect the dynamics of large tick stocks and leave unaltered the small tick stocks. It is also know that fees/rebates are slightly higher on NASDAQ than on NYSE.  We noticed that the main difference actually lies in the trade size, which appear to be on average  smaller on NASDAQ than on NYSE for the large tick stocks (in some sense, one could say that the large ticks on NASDAQ have a small tick behaviour, consistent with the possibility of having trades within the spread). 

The latter point in fact suggests that the average trade size should also be taken into consideration when it comes to the quest of a universal market invariant. Quite remarkably, the two-branch structure nearly disappears when $\langle I\rangle$ is plotted against average trade cost ${\cal C}=\langle S Q\rangle$, in addition to revealing a roughly linear dependence (see top left inset, and the last section below for a quantitative interpretation). Moreover, the average rescaled invariant equals $0.86\pm 0.54$, where the uncertainty reflects the root mean square  cross-sectional dispersion,  is now of the same order than the corresponding value for futures ($ 0.30\pm0.09$, see Fig.~\ref{NewFig}).
As was the case for futures contracts, the distributions of $I$ have power law tails, with tail exponents fluctuating around $\mu \approx 4$, but with no particular tick dependence. The values of $\mu$ for twelve randomly chosen stocks can be found in Tab.~\ref{SummaryStocks}.

\section{3. Theoretical Analysis} 
\label{secAnalysisI}

We now turn to a theoretical analysis of the above results, with the aim of gaining a better understanding of the 3/2 scaling law, observed both on futures 
contracts and on single stocks. In most of this section, we redefine the trading activity as $\widetilde W=V\sigma$, without the price $P$ which is irrelevant for the 
points we want to make. The role of the price will be discussed in the next section. 

We first propose a very simple argument that suggests to decompose the $N$-dependence of the trading activity $\widetilde W$ into two parts, one coming from the $N$-dependence of $\sigma$ and the other coming from the $N$-dependence of $V$. This decomposition reveals a much more subtle picture, where neither $\sigma$ nor $V$ behave as naively expected, but the product of the two indeed scales approximately as $N^{3/2}$. Most of this section is about futures, for which the story is
surprisingly complex, whereas stocks behave more trivially and are discussed at the end. 

\subsection{A naive argument}

If one assumes that there is a well-defined average trading frequency $\phi$ (defined as the number of trades per unit time) and a well-defined average trade size
$Q_0$, then after time $\tau$ one expects the following two relations:
\be
N = \phi \times \tau; \qquad V = Q_0 \times N.
\ee
Since the (log)-price is close to a random walk, one should also have:
\be
\sigma = \varsigma_0 \sqrt{\tau} :=  \frac{\varsigma_0}{\sqrt{\phi}} \sqrt{N} , 
\ee
where $\varsigma_0$ is a constant which, according to Ref. \cite{Wyart2008}, is proportional to the spread $S$. Hence, 
\be \label{eqQS}
\widetilde W =  V \sigma =  \frac{\varsigma_0 Q_0}{\sqrt{\phi}}  N^{3/2} \propto Q_0 S N^{3/2}.
\ee
This appears to fully explain the 3/2 scaling law, which would then be an almost trivial observation.  
Although this will indeed turn out to be the correct mechanism for individual stocks, futures contracts reveal a much more intricate story, 
at least for large ticks and small time intervals $\tau$ -- more precisely when the volatility on scale $\tau$ is small compared to the tick size $s$.

\subsection{A more complex picture} 
\label{complex}
\begin{figure}[t!]
\centering 
\includegraphics[scale=0.52]{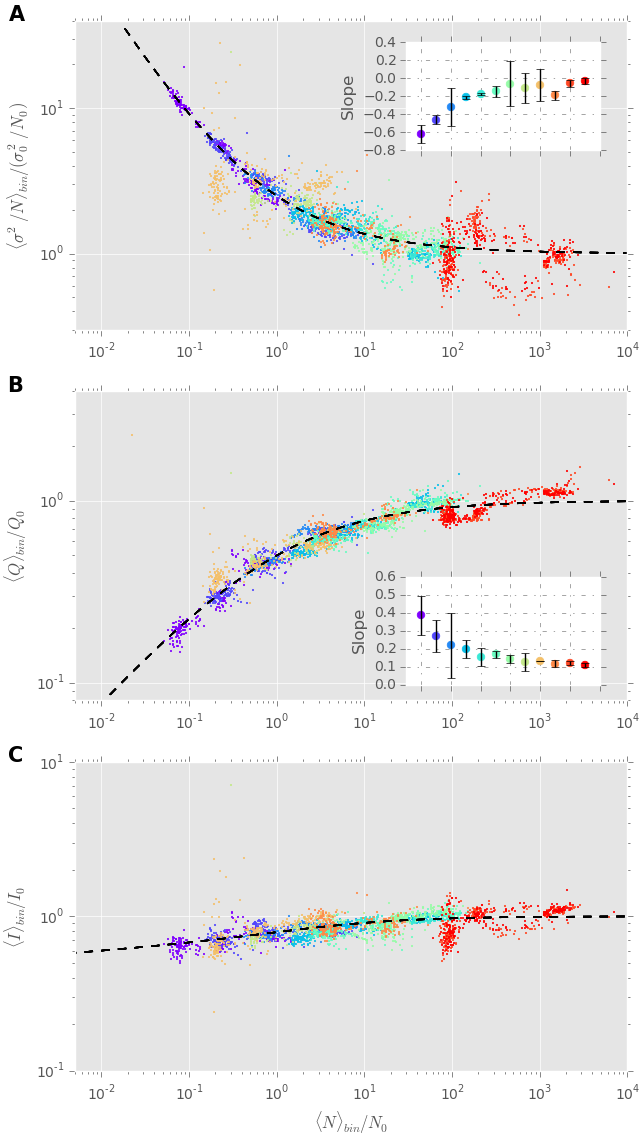}
 \caption{Data for {twelve} futures contracts at high frequency (5 minute bins). (A) Rescaled signature plot obtained by fitting {$\langle \sigma^2/N\rangle_{\text{bin}}$ against $\langle N\rangle_{\text{bin}}$} as given by Eq.~\eqref{sigma}, with $a=0.5$. (B) Rescaled average trade size obtained by fitting Eq.~\eqref{Q} with $\nu = 0.54$ to the data, as a function of { $\langle N\rangle_{\text{bin}}/N_0$. (C) Rescaled plot of  $\langle \tilde I\rangle_{\text{bin}}/I_0$ against $\langle N\rangle_{\text{bin}}/N_0$, resulting from (A) and (B), and consistent with Eq.~\eqref{I}. The values of  $N_0$, $\sigma_0$ and $Q_0$ are reported in Tab.~\ref{RescFutTable}. The insets of plots (A) and (B) display the slopes obtained from linear regression of $\langle \sigma^2/N\rangle_{\text{bin}}$ and $\langle Q\rangle_{\text{bin}}$ against $\langle N\rangle_{\text{bin}}$} respectively for each of the {twelve} futures contracts at hand.}
\label{RescFutAsset}
\end{figure}

We first analyze independently the above two scaling laws ($\sigma \sim \sqrt{N}$, $V \sim N$) on our pool of { twelve} futures contracts. We focus first on $\tau$ = 5-minute bins, 
a good trade-off between high frequency and noise.  The insets of Figs~\ref{RescFutAsset}(A) and (B) show the exponents 
obtained by a power-law fit of  $\langle \sigma^2/N\rangle_{\text{bin}}$ \textit{vs.} $\langle N \rangle_{\text{bin}}$ (the so-called signature plot) and $\langle Q \rangle_{\text{bin}}$ \textit{vs.} $\langle N \rangle_{\text{bin}}$ respectively. 
As can be seen in these figures, small tick futures are indeed consistent with the expected $\sigma \sim \sqrt{N}$ and $V \sim N$ behaviour. For the 
large tick futures one rather finds  $\sigma \sim N^{\beta}$ and $V \sim N^{\gamma}$, with $\beta < 1/2$ and $\gamma > 1$, suggesting of (i) 
a sub-diffusive price dynamics and (ii) an effective average trade size that increases with $N$. The rather puzzling fact, however, is the two exponents
appear to conspire to give $\beta +\gamma \approx 3/2$ such that the scaling $\widetilde W \sim N^{3/2}$ indeed holds regardless of tick size. 

\subsection{Delayed diffusion for large ticks}

A sub-diffusive behaviour for large tick contracts is in fact expected at short times, because a continuous random walk $B(\tau)$ that is constrained to
take integer values $[B(\tau)]=n \times s$ (where $n$ is an integer and $s$ the tick size) can easily be shown to fluctuate as $\tau^{1/4}$ when $\tau$ is 
small (instead of the usual $\sqrt{\tau}$ behaviour). Furthermore, one expects a large amount of microstructural high frequency noise on the price when 
the tick is large. A simple way to account for these two effects is to postulate the following effective diffusion law: \footnote{For an inspiring approach of the price dynamics of large tick assets, 
see a recent paper by Dayri and Rosenbaum \cite{Dayri2013}. Their analysis might shed light on the results discussed here.}
\begin{eqnarray}
 \sigma &=& \sigma_0  \left[ a + \left(\frac{N}{N_0}\right)^{\frac 12} + \frac{N}{N_0} \right]^{\frac 12} \ ,
\label{sigma}
\end{eqnarray}
where $a$ accounts for the high-frequency noise and $N_0$ is a characteristic number of trades such that the usual random walk behaviour is 
expected for $N \gg N_0$. One expects that $N=N_0$ roughly corresponds to a one tick move, so that $\sigma_0$ should be of order $s$ and, correspondingly, 
$a$ of order unity (since the amount of microstructural noise should be set by the tick size). The parameters $N_0$ and $\sigma_0$ are to be fitted to the data for each contract. 
We will see below that these expectations are indeed confirmed 
by the data (see Tab.~\ref{RescFutTable}). 
Note that for large ticks and small trade sizes, one has $N_0 \gg 1$ and a very wide region where the anomalous sub-diffusion law $N^{1/4}$ 
holds. In the other limit $N_0 \lesssim 1$, the diffusive regime is almost immediately reached.

\subsection{Master curves for volatility and volumes}

Now, as shown in Figure~\ref{RescFutAsset}(A), the signature plots of {\it all} our futures contracts can be {quite} convincingly rescaled on a unique master curve 
given by Eq. (\ref{sigma}), with appropriately chosen values of $\sigma_0$ and $N_0$ that are reported in Tab.~\ref{RescFutTable}. We fixed  $a=0.5$ for 
all contracts, consistent with an overall goodness of fit when considering the {twelve} futures together. As expected, $\sigma_0$ is indeed found to be of the 
order of the tick size.

We now turn to the effective trade size $Q=V/N$, which can be similarly rescaled on a unique master curve by the following formula -- see Fig.~\ref{RescFutAsset}(B):
\begin{eqnarray}
 Q &=& Q_0   \left[ 1 +\left(\frac{N}{N_0}\right)^{-\nu}\right]^{-1} \ ,
\label{Q}
\end{eqnarray}
where the value of $N_0$ is fixed, contract by contract, to the very value favored by the rescaling of the signature plot. The only free parameters are $Q_0$  (reported  in Tab.~\ref{RescFutTable}), and the exponent $\nu \approx 0.54$, determined by the minimization of the overall error function of the aggregated data from all the futures. Note that $Q_0$ is the asymptotic value (for large $N$) of the average volume per trade. Figure \ref{Q0} displays $Q_0$ against the average volume at the bid/ask $V_{\text {best}} = (V_{\text {bid}}+V_{\text {ask}})/2$. As expected $Q_0 \sim V_{\text {best}}$ for small tick stocks, but grows sub-linearly for large tick stocks where trades only represent 
a smaller and smaller fraction of the available volume. 

\begin{figure}[t!]
  \centering 
  \includegraphics[scale=0.54]{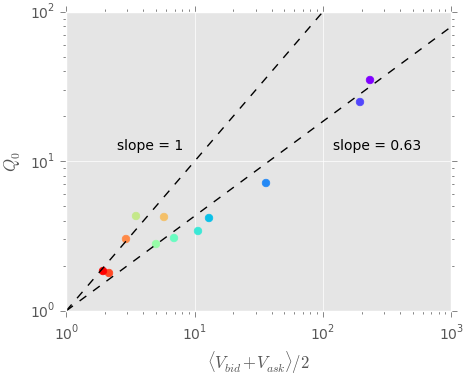}
 \caption{Plot of $Q_0$ - as obtained from fitting Eq.~\eqref{Q} to the data -  as a function of average volume at the bid/ask ({see Tab.~\ref{RescFutTable}}), for {twelve} futures contracts.}
\label{Q0}
\end{figure}

\subsection{Deviations from the 3/2-law}

\begin{figure}[b!]
  \centering 
  \includegraphics[scale=0.52]{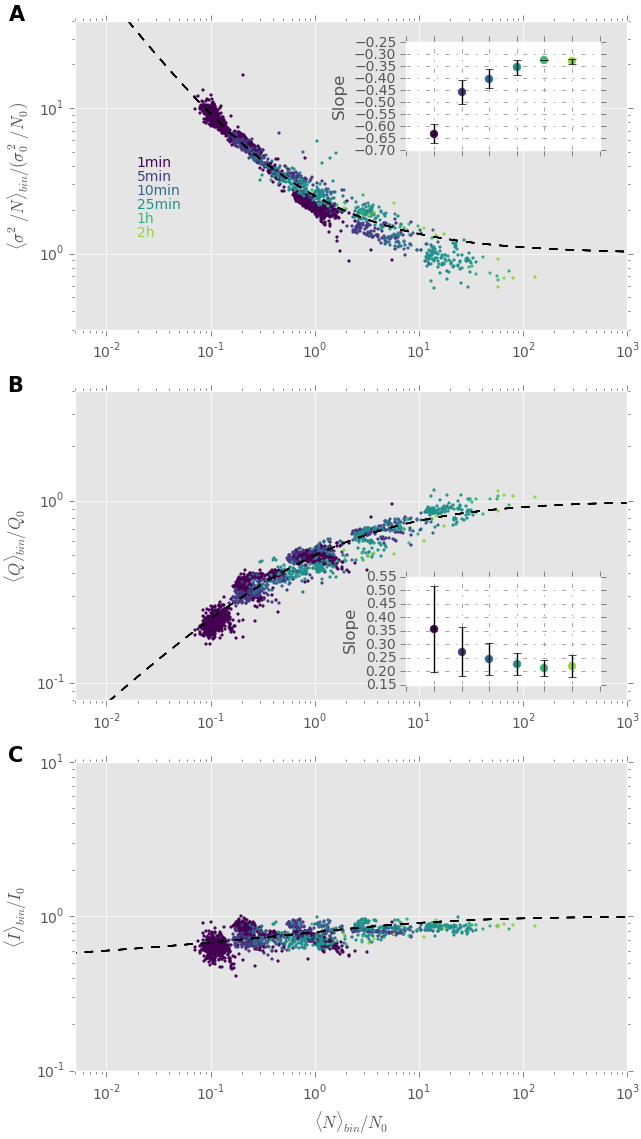}
 \caption{Figure analogous to Fig.~\ref{RescFutAsset}, only for the SPMINI futures contract at different sampling frequencies: $\tau=$ 1 min, 5 min, 10 min, 25 min, 1h and 2h bins. The value of $N_0$ has been set 
 to that measured on 5 minute bins. The values of $\sigma_0$ and $Q_0$ are left free but are found to be roughly constant across sampling frequencies (see Tab.~\ref{RescFutTableTime}). The mild increase of $I$ with $\tau$ (bottom
 graph) should be compared to the results shown in Fig.~\ref{NewFig} .}
\label{RescFutTime}
\end{figure}

We have shown that the deviations from simple diffusion and naive additivity of trade sizes can be rationalized by two more sophisticated 
scaling laws, Eqs.~\eqref{sigma} and \eqref{Q}, leading to two universal master curves. Combining these two laws and letting $n=N/N_0$ and $I_0 = Q_0\sigma_0/\sqrt{N_0}$, allows one to write: 
\begin{eqnarray}
\frac{\widetilde W}{N^{3/2}} = \tilde I = I_0 \, \frac{\left( a n^{-1} + n^{-1/2} + 1 \right)^{1/2}}{1 +n^{-\nu}} \ .
 \label{I}
\end{eqnarray}
Equation \eqref{I} offers a quantitative unifying picture of the above observation that $\beta+\gamma \approx 3/2$ regardless of tick size. Note that for $N \gg N_0$, $\tilde I \rightarrow I_0$ (such that $I_0$ can be seen as the asymptotic trading invariant at large $N$), while for $N\ll N_0$, $\tilde I \rightarrow I_0 a \,n^{\nu-1/2}$ which is also nearly constant when $\nu \approx 1/2$. Therefore, our scaling analysis suggests that $\widetilde WN^{-3/2}$ has in fact a residual $N$ dependence.  Figure~\ref{RescFutAsset}(C) displays a plot of $\langle \tilde I\rangle_{\text{bin}}/I_0$ against $\langle N\rangle_{\text{bin}}/N_0$, showing that the data can be rescaled onto a single master curve, as given by Eq.~\eqref{I}, and indeed revealing a small, but significant variation with $N$.

\subsection{Time rescaling}

Eq. (\ref{sigma}) describes the crossover between a sub-diffusive regime for small $N$ and a purely diffusive regime at large $N$, and was calibrated on 
different contracts for the same {bin size} $\tau=$ 5 min. If our line of reasoning is correct, the {\it very same rescaling} should hold when focusing on a given contract 
but letting $\tau$ vary, in such a way that $N/N_0$ itself increases. This assumption is indeed in agreement with the data on all futures contracts. 
For the sake of clarity, we only show data for the SPMINI contract -- but other contracts behave similarly. We considered $\tau=$ 1 min, 5 min, 10 min, 25 min, 1h and 2h, and set $N_0$ to the value obtained in the previous paragraph for 5 minute bins. Figure~\ref{RescFutTime} displays plots analogous to those of Fig.~\ref{RescFutAsset}, but for the SPMINI contract across the different sampling frequencies. As one can see, our extend scaling hypothesis allows one to explain both the variations across 
contracts for a given $\tau$ and across time intervals.  The scaling exponents of $\sigma(N)$ and $V(N)$ do converge to their natural values ($1/2$ and $1$) as  $N$ becomes much larger than $N_0$. However, the fact that Eqs.~\eqref{sigma} and \eqref{Q} hold for all $\tau$ explains why $\tilde I/I_0$ varies mildly with $\tau$, as Fig.~\ref{NewFig} above also demonstrated.

\subsection{Naive scaling for single stocks}

We now test the naive scalings $\sigma \sim \sqrt{N}$ and $V \sim N$ for {$\tau = 5$ min} by regressing  $\log \sigma$ and $\log Q$ \textit{vs.} $ \log N$ for each of the three hundred stocks 
individually. Keeping with the notations introduced above, we find that the slopes $\beta$ and $\gamma$ of these regressions show no significant systematic dependence on the tick size. A cross sectional determination of these two exponents yields {$\beta= 0.51 \pm 0.06$ and $\gamma = 1.04 \pm 0.07$}, where the uncertainty {again} reflects the root mean square  cross-sectional dispersion. At the daily time-scale one has equivalently {$\beta= 0.54 \pm 0.10$ and $\gamma = 1.02 \pm 0.12$}. Therefore, for all time scales $\tau \geq 5$ min, one can assume that the natural asymptotic scaling holds for all stocks, which trivially leads to the 3/2-law.

Still, it is surprising that the deviation from $\sigma \sim \sqrt{N}$, clearly observed for futures, does not seem to be present for stocks. In order to understand
this difference, we display in Fig.~\ref{Dif_Fut_Sto}  the scatter plot $\sigma^2/N$ vs. $N$ for both our largest tick future (TBOND) and our largest tick stock ({Applied Materials Inc}, {see Tab.~\ref{SummaryStocks}}). The black line represents an average on consecutive log-spaced bins. As one can see, while for the future the average slope of the black line is consistent with the sub-diffusive behaviour discussed above, this is not the case for the stock which is on average rather flat in the region where most points are found. 

This effect can actually be attributed to the fact that futures are traded on three different time zones, so that $5$-minute bins where trading is slow (i.e. $N$
small compared to $N_0$) are much more represented in the data than for stocks. The latter are indeed only active on American regular trading hours for which periods of very low activity are much rarer \footnote{{Note that we only have 5-minute binned data for stocks, preventing us to zoom on shorter time intervals $\tau$ for which the sub-diffusive behaviour for large ticks should eventually show up.}}. Therefore the regression slope $\beta$ for futures is expected to be more sensitive to sub-diffusive effects. Furthermore, the 
volatility of single stocks is a factor $2-4$ smaller than the volatility of large tick futures, meaning that for the same relative tick size, discretisation effects are expected to be smaller for stocks. In this sense, large tick futures have a ``larger tick'' than large tick stocks! 

\begin{figure}[t!]
\centering 
\includegraphics[scale=0.5]{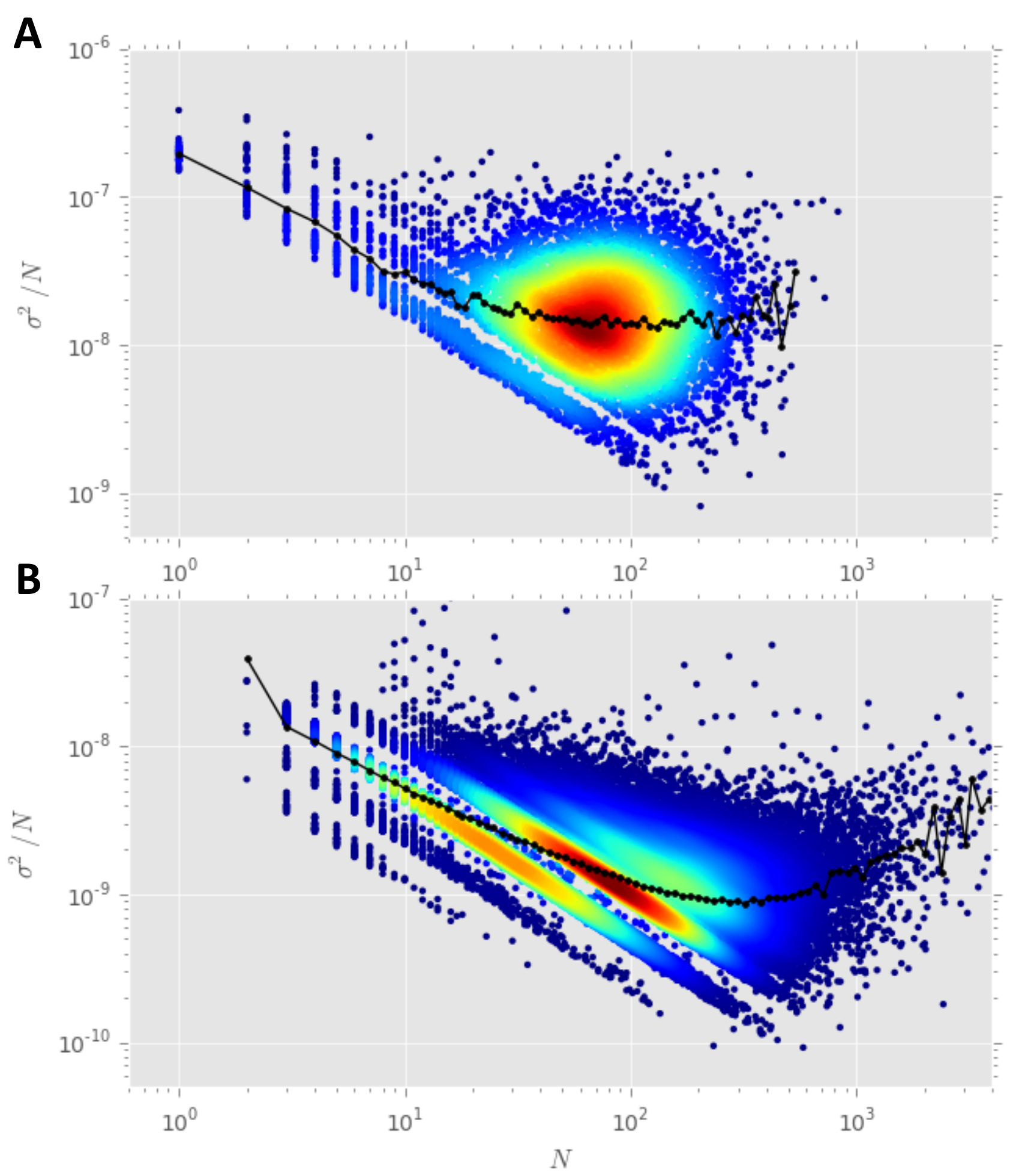}
 \caption{Signature scatter plot for ({A}) our largest tick stock ({Applied Materials Inc}, {see Tab.~\ref{SummaryStocks}}) and ({B}) our largest tick future (TBOND) computed with all 5-minute bin data. 
 The black line represents an average on consecutive log-spaced bins, and the color code indicates the density of data. This graph shows that
 discretisation effects are much larger for the TBOND than for {Applied Materials Inc}.}
\label{Dif_Fut_Sto}
\end{figure}

\section{4. Prices, spreads and a new definition of the trading invariant} 
\label{secSpread}

As noted in Fig.~\ref{NewFig} and the top left inset of Fig.~\ref{ProbaStocks}, the quantity  $I/{\cal C}$ where ${\cal C}=\langle SQ\rangle $ appears to be more universal across assets than $I$ itself, both for futures and for stocks. The microstructural quantity $\langle SQ\rangle$, where $S$ and $Q$ respectively denote the spread (in dollars per share) and trade size (in shares), corresponds to the average cost of trading. As a matter of fact, there is a direct element supporting that the spread should be included in the discussion. Indeed, the theory presented in Wyart~\textit{et~al.} \cite{Wyart2008}, based on zero marginal profits for market makers, {see also \cite{MRR}}, predicts that for small tick {contracts}, $\sigma(N) = c S \sqrt{N}$, where $S$ is the spread and $c$ a numerical constant of order unity. Such a prediction was found to be very accurately obeyed by data, see Ref. \cite{Wyart2008}.  These observations naturally lead to a slightly amended definition of the trading invariant, that has the additional virtue of leading to a unit-less quantity, at variance with KO's definition where the invariant has dollar units. Our proposal, inspired from Eq. (\ref{eqQS}), is thus to consider the quantity ${\cal I}$, defined as:
\be
{\cal I} = \frac{PV \sigma}{{\cal C} N^{3/2}} \ , \qquad {\cal C} = \langle SQ \rangle
\ee}
where both the price and the spread are expressed in dollars per share, and ${\cal C}$ is the average spread cost, a concept actually at the very basis of the original trading invariant proposed by Kyle and co-authors.
The quantity ${\cal I}$ is clearly less scattered across contracts than $I$ itself. Actually, ${\cal I}$ is of order unity for both individual stocks and futures, which is quite remarkable in view of the strong differences between these asset classes. In any case, we find it much more convincing to define a unit-less quantity as a plausible candidate for a genuine market invariant or quasi-invariant, in a sense we discuss now. 

\section{5. Conclusion}

Let us summarize what we have achieved in this work:
\begin{itemize}
\item The most important result, to our eyes, is  the 3/2-law, stating that the amount of risk $W$ exchanged in markets (i.e. price times volume times volatility) scales like the 3/2 power of the number of trades $N$. We have shown that this holds very precisely across all 12 futures contracts and 300 single stocks, and across all times scales $\tau$, thereby considerably extending the results obtained by Andersen et al. \cite{Andersen2015} on the E-mini S\&P futures. 
\item The second result is that the trading invariant $I=W/N^{3/2}$ proposed by Kyle and Obizhaeva is in fact quite different 
for different contracts, in particular between futures and single stocks. Furthermore, this quantity has dollar units, which makes its invariance property dubious. 
On the basis of a combination of dimensional, theoretical and empirical arguments, we have proposed that a more natural candidate should rather be  ${\cal I}=I/{\cal C}$, where $\cal C$ is the 
average (spread) cost of a trade. In fact, this rescaling is in line with Kyle and Obizhaeva's initial intuition that $I$ should be related to the cost of a ``bet''. 
Whether the weak residual dependence of $\cal{I}$ on the tick size is real or comes from some spurious biases is left for future investigations.
\item Third, we have unveiled two remarkable master curves for the volatility $\sigma$ and the traded volume $V$ as a function of $N$, in the case of large tick futures contracts.  We have argued for the existence of a characteristic number of trades $N_0$ above which the naively expected behaviour $\sigma \sim \sqrt{N}$ and $V \sim N$ hold, but below which strong deviations appear, induced by the size of the tick. 
\end{itemize}
A synthetic way to summarize all our findings is to generalize and amend the dimensional analysis formula, Eq. (\ref{general}), as:
\be\label{improved}
\frac{PQ}{\cal C} = f_{\text{asset}}\left(\frac{Q \sigma^2}{V}, \frac{P \sigma \sqrt{\tau}}{s}\right),
\ee
where the left-hand side is dimensionless. The function $f_{\text{asset}}$ now depends on the asset class (futures {\it vs.} stocks) as well as on a second dimensionless argument which involves the time interval $\tau$ and the tick size $s$. For large enough $\tau$, or small enough $s$, one expects this dependence to disappear, i.e. $f(x,y \to \infty) = x^{-1/2}$ leading back to Kyle and Obizhaeva's hypothesis (up to the presence of the spread $S$, {rather than $I$,} in the left hand side). In the other limit, $f(x, y \to 0)$,  could in principle behave very differently, but our detailed analysis above has revealed that $f(x,y)$ {remains} close to $x^{-1/2}$, with only a weak dependence on the second argument -- see again Fig.~\ref{RescFutAsset}(C). In other words, the 3/2-law holds much beyond the regime 
$y \gg 1$, where it is expected on the basis of naive scaling. We do not have, at this stage, a detailed understanding of whether this is merely coincidental, or whether there is a deeper principle enforcing this property. We leave this as an open question for future work.

\subsection{Acknowledgments} 

We wish to thank F. Patzelt for helping with the data analysis, as well as S. Hardiman, I. Mastromatteo, { L. Duchayne}, Z. Eisler, J. Kockelkoren, M. Potters, M. Vladkov, A. Darmon 
and C.-A. Lehalle for very fruitful discussions. We also thank the referee for making a crucial suggestion that helped us hone in on our final definition of the trading invariant ${\cal I}=I/{\cal C}$.


\bibliographystyle{h-physrev}
\bibliography{Bibli_Benzaquen2016}

\begin{table*}[t!]
  \centering
\begin{tabular}{llrrrrrrrrrrr}
\toprule
     Name &$\langle$Spread$\rangle$&    $\langle N\rangle$ &   $\langle Q\rangle$   &     $\langle V\rangle$ &       $\langle P\rangle$ &  $\langle \sigma\rangle$ &   $\langle V_{\text{bid}}\rangle$ &   $\langle V_{\text{ask}}\rangle$ &  $\alpha$ &      $\langle I\rangle$ &    $\mu$ \\
\toprule
      TBOND &             1.007 &  18 &  10.8 &  191 &  140960 &   1.22 &  206.1 &  206.5 &   1.58 &  98.0 &  2.49 \\
     SPMINI &             1.011 &  45 &  11.2 &  499 &   81025 &   1.46 &  183.8 &  185.4 &   1.50 &  30.5 &  2.51 \\
        EUR &             1.021 &  22 &   4.1 &   89 &  163831 &   1.70 &   33.9 &   33.7 &   1.52 &  18.6 &  2.26 \\
   NSDQMINI &             1.120 &  17 &   2.8 &   47 &   58969 &   2.04 &   11.9 &   12.0 &   1.56 &   7.0 &  2.18 \\
     DJMINI &             1.157 &  13 &   2.5 &   33 &   73777 &   1.98 &    9.8 &    9.8 &   1.53 &   6.7 &  2.33 \\
     CRUDE0 &             1.227 &  21 &   2.2 &   48 &   94290 &   2.50 &    6.4 &    6.4 &   1.55 &  12.0 &  1.37 \\
      GOLD0 &             1.323 &  21 &   2.2 &   45 &  153092 &   2.61 &    4.6 &    4.6 &   1.58 &  12.5 &  1.63 \\
   NCOFFEE0 &             2.066 &   6 &   2.0 &   12 &   59407 &   1.57 &    3.1 &    3.0 &   1.66 &  24.7 &  1.52 \\
  WTICRUDE0 &             2.258 &   9 &   1.8 &   17 &   94520 &   2.71 &    5.2 &    5.1 &   1.52 &  16.5 &  1.41 \\
   NCOTTON0 &             3.628 &   5 &   2.0 &   10 &   40375 &   2.38 &    2.5 &    2.5 &   1.47 &  10.9 &  1.64 \\
   HEATOIL0 &             4.950 &   9 &   1.7 &   15 &  122626 &   6.56 &    2.0 &    2.0 &   1.48 &  15.6 &  1.62 \\
   RBGASOL0 &             6.052 &   9 &   1.7 &   15 &  116292 &   7.49 &    1.8 &    1.8 &   1.46 &  18.1 &  1.61 \\
\toprule
  \end{tabular}
  \caption{Summary table for twelve futures contracts. {Values are computed in one minute bins}.  The average spread and volatility are given in units of tick, the trade size is given in number of contracts, the volume is given in contracts per unit time.  The ``invariant'' $I=PV\sigma/N^{3/2}$ is given in dollars. {{All averages are defined as $\langle X \rangle := \exp\left[ \mathbb E (\log X) \right]$.}}}
\label{SummaryFut}
\end{table*}

\begin{table*}[t!]
  \centering
\begin{tabular}{llrrrrrrrrr}
\toprule
     Name &$\langle$Spread$\rangle$&    $\langle N\rangle$ &   $\langle Q\rangle$   &     $\langle V\rangle$ &       $\langle P\rangle$ &  $\langle \sigma\rangle$ &    $\alpha$ &      $\langle I\rangle$ &    $\mu$ \\
\toprule
  YHOO\_UQ &             1.006 &   68 &  325.2 &  22025 &  16.0 &   1.28 &   1.55 &  0.51 &  3.67 \\
  EBAY\_UQ &             1.097 &  124 &  153.4 &  19045 &  41.9 &   4.00 &   1.58 &  0.55 &  4.26 \\
   WMI\_UN &             1.099 &   21 &  249.3 &   5146 &  33.6 &   2.05 &   1.53 &  1.13 &  3.56 \\
    PG\_UN &             1.172 &   43 &  280.9 &  12057 &  66.1 &   3.18 &   1.52 &  1.37 &  3.42 \\
    FE\_UN &             1.335 &   21 &  187.8 &   3888 &  44.9 &   2.60 &   1.50 &  1.07 &  3.98 \\
  FAST\_UQ &             1.574 &   35 &  117.4 &   4133 &  44.9 &   3.66 &   1.53 &  0.72 &  4.17 \\
 DISCA\_UQ &             1.923 &   26 &  111.5 &   2853 &  52.5 &   3.45 &   1.53 &  0.76 &  3.89 \\
   ZMH\_UN &             2.204 &   17 &  148.0 &   2490 &  62.6 &   3.72 &   1.58 &  1.34 &  3.70 \\
   MON\_UN &             2.628 &   30 &  153.5 &   4577 &  82.7 &   6.06 &   1.54 &  1.70 &  3.95 \\
   MGA\_UN &             3.319 &   11 &  132.1 &   1508 &  43.1 &   3.63 &   1.53 &  1.42 &  3.58 \\
   ALXN\_UQ &             5.568 &   24 &  110.5 &   2628 &  93.5 &   7.85 &   1.54 &  1.78 &  3.47 \\
   MARY\_UQ &             7.613 &   13 &  126.9 &   1679 &  58.1 &   6.82 &   1.56 &  2.38 &  3.57 \\
\hline
   {AMAT\_UQ }&             1.003 &   55 &  338.6 &  18535 &  11.4 &   1.10 &   1.49 &  0.51 &  3.20 \\
 \toprule
  \end{tabular}
  \caption{Summary table for a random subset of twelve stocks and Applied Materials Inc.~(used in Fig.~\ref{Dif_Fut_Sto}(A)). Values are computed in five minute bins. Units are identical to those of Tab.~\ref{SummaryFut}. {{All averages are defined as $\langle X \rangle := \exp\left[ \mathbb E (\log X) \right]$.}}}
  \label{SummaryStocks}
\end{table*}

\clearpage

\begin{table}[t!]
  \centering
\begin{tabular}{llrrrrr}
\toprule
      Name &$\langle$Spread$\rangle$&      $N_0$ &  $\sigma_0$  &   $Q_0$ & $I_0$ & $\langle V_{\text {best}} \rangle$ \\
\toprule
      TBOND &             1.008 &  177.63 &    1.23 &  35.18 &  3.25 &    224.6 \\
     SPMINI &             1.012 &  156.88 &    1.34 &  24.77 &  2.65 &    190.3 \\
        EUR &             1.022 &   40.06 &    1.26 &   7.18 &  1.43 &     35.3 \\
   NSDQMINI &             1.130 &    8.07 &    0.98 &   4.16 &  1.44 &     12.7\\ 
     DJMINI &             1.171 &    4.71 &    0.84 &   3.40 &  1.31 &     10.4 \\
          CRUDE0 &             1.242 &    5.85 &    1.08 &   3.06 &  1.36 &      6.8\\
                GOLD0 &             1.340 &    4.46 &    0.99 &   2.79 &  1.31 &      4.9\\ 
   NCOFFEE0 &             2.143 &   20.16 &    1.87 &   4.32 &  1.80 &      3.3 \\
  WTICRUDE0 &             2.304 &   45.17 &    2.78 &   4.24 &  1.75 &      5.6  \\
   NCOTTON0 &             3.819 &    2.16 &    1.25 &   3.01 &  2.56 &      2.8    \\
   HEATOIL0 &             5.092 &    0.08 &    0.51 &   1.80 &  3.28 &      2.1      \\
   RBGASOL0 &             6.253 &    0.07 &    0.55 &   1.86 &  3.78 &      1.9 \\
\toprule
\end{tabular}
  \caption{Values of $N_0$, $\sigma_0$ {(in ticks)}, $Q_0$  obtained by fitting the data of { twelve} futures ({with $\tau = 5$  min}) to Eqs.~\eqref{sigma} and \eqref{Q} (see Fig.~\ref{RescFutAsset}), $I_0 = Q_0\sigma_0/\sqrt{N_0}$ as well as $V_{\text {best}} = (V_{\text {bid}}+V_{\text {ask}})/2$.  Averages  are defined as $\langle X \rangle := \exp\left[ \mathbb E (\log X) \right]$.}
  \label{RescFutTable}
\end{table}

%
%
%

\bigskip

\begin{table}[h!]
  \centering
\begin{tabular}{llrrrrr}
\toprule
 Binsize &      $\sigma_0$ &    $ Q_0$  &$I_0$\\
\toprule
 1min &            1.21 &  33.80 &  3.27 \\
 5min &            1.34 &  24.77 &  2.65 \\
10min &            1.45 &  21.98 &  2.54 \\
25min &            1.64 &  19.14 &  2.51 \\
   1h &            1.83 &  16.93 &  2.47 \\
   2h &            1.94 &  15.91 &  2.46 \\

\toprule
\end{tabular}
  \caption{Values of $\sigma_0$ {(in ticks)} and $Q_0$ obtained by fitting the data of  the SPMINI at different sampling frequencies to Eqs.~\eqref{sigma} and \eqref{Q} (see Fig.~\ref{RescFutTime}), {and $I_0 = Q_0\sigma_0/\sqrt{N_0}$}.\bigskip\bigskip\bigskip\bigskip}
  \label{RescFutTableTime}
\end{table}

\end{document}